\documentclass[english,aps,prl,showpacs,amsmath,amssymb,twocolumn]{revtex4}
\usepackage{graphicx}

\makeatletter

\usepackage{dcolumn}

\usepackage{bm}

\usepackage{epsf}

\usepackage{babel}
\makeatother
\begin{document}

\title{The predictability of letters in written english}

\author{Thomas Sch\"urmann and Peter Grassberger}

\affiliation{Department of Theoretical Physics, University of Wuppertal, Germany}

\begin{abstract}
We show that the predictability of letters in written English texts depends strongly on their position in the word. The first letters are usually the least easy to predict. This agrees with the intuitive notion that words are well defined subunits in written languages, with much weaker correlations across these units than within them. It implies that the average entropy of a letter deep inside a word is roughly 4-5 times smaller than the entropy of the first letter.
\end{abstract}

\pacs{89.70+c, 02.50.Fz, 05.45.Tp}

\maketitle
Since language is used to transmit information, one of its most quantitative characteristics is the entropy, i.e., the average amount of information (usually measured in bits) per character.

Entropy as a measure of information was introduced by Shannon 
\cite{SW}. He also performed extensive experiments \cite{S} using the ability of humans to predict continuations of printed text. This and similar experiments \cite{CK,LR} led to estimates of typically $\approx 1-1.5$ bits per character.

In contrast, the best computer algorithms whose prediction is based on sophisticated statistical methods reach entropies of $\approx 2-2.4$ bits \cite{BCW}. Even this is better than what commercial text compression packages achieve: starting from texts where each character is represented by one byte, they typically achieve compression ratios $\approx 2$, corresponding to $\approx 4$ bits/character. These differences result from different abilities to take into account long-range correlations which are present in all texts and whose utilization requires not only a good understanding of language but also substantial computational resources. 

Formally, Shannon entropy $h$ of a letter sequence $(...,s_{-1},s_0,s_1,...)$ over an alphabet of $d$ letters is given by 

\begin{eqnarray}\label{eq2}
h=&-&\lim\limits_{n\to\infty}\sum_{s_{-n},...,s_0}
p(s_{-n},...,s_0)\\
&\times&\log\, p(s_0|s_{-1},...,s_{-n})\nonumber\\
&=&\lim\limits_{n\to\infty}\langle -\log\, p(s_0|s_{-1},...,s_{-n})\rangle
\end{eqnarray}

where $p(s_{-n},...,s_0)$ is the probability for the letters at position $-n$ to $0$ to be $s_{-n}$ to $s_0$, and $p(s_0|s_{-1},...,s_{-n})=\frac{p(s_{-n},...,s_0)}{p(s_{-n},...,s_{-1})}$. The second line of this equation tells us that $h$ can be considered as an average over the {\it information} of {\it bit number}. While Eq. (\ref{eq2}) obviously assumed stationarity, we can define the latter also for nonstationary sequences, provided they are distributed according to some probability $p$ which satisfies the Kolmogorov consistency conditions. The information of the $k$th letter when it follows the string $...,s_{k-2},s_{k-1}$ is thus defined as:

\begin{eqnarray}\label{bitnum}
\eta_k=\lim\limits_{n\to\infty} \log \frac{1}{p(s_k|s_{k-1},...,s_{k-n})}
\end{eqnarray}

Notice that this depends both on the previous letters (or "contexts" \cite{WRF}) and on $s_k$ itself. If the sequence is only one-sided infinite (as for written texts), we extend it to the left with some arbitrary but fixed sequence, in order to make the limes in Eq. (\ref{bitnum}) well defined.  

When trying to evaluate $\eta_k$, the main problem is the fact that $p(s_k|s_{k-1},...,s_{k-n})$ is not known. The best we can do is to obtain an estimator $\hat{p}(s_k|s_{k-1},...,s_{k-n})$ which then leads to an information estimate $\hat{\eta}_k$, and to:

\begin{eqnarray}\label{eq4}
\hat{h}_N=\frac{1}{N}\sum\limits_{k=1}^N\eta_k
\end{eqnarray}

for a text of length $N$. This can be used also for testing the quality of the predictor $\hat{p}(s_k|s_{k-1},...,s_{k-n})$: the best predictor is that which leads to the smallest $\hat{h}$. This is indeed the main criterion by which $\hat{p}(s_k|s_{k-1},...,s_{k-n})$ is constructed.

In this way we do not only get an estimate $\hat h$ of $h$, but we can investigate the predictability of individual letters within specific contexts. The fact that different letters have different predictabilities is of course well known. If no contexts are taken into account at all, then the best predictor is based on the frequencies of letters, making the most frequent ones the easiest to predict. Studies of these frequencies exist for all important languages.

Much less effort has gone into the context dependence. Of course, the next natural distribution after the single-letter probabilities are the distributions of pairs and triples which give contexts of length 1 and 2, and which have also been studied in detail \cite{BCW}. But these distributions do not directly reflect some of the most prominent features of written languages, namely, that they are composed of subunits (words, phrases) which are put together according to grammatical rules. 

In the following, we shall study the simples consequences of this structure. If words are indeed natural units, it should be much easier to predict letters coming late in the word - where we have already seen several letters with which they should be strongly correlated - than letters at the beginnings of words. Surprisingly, this effect has not jet been studied in the literature, maybe due to a lack of efficient estimators of entropies of individual letters. A similar, but maybe less pronounced effect is expected with words replaced by phrases. 

In our investigation, we use an estimator which is based on minimizing $\hat h$. Technically, it builds a rooted tree with contexts represented as path starting at some inner node and ending at the root. The tree is constructed such that each leaf corresponds to a context which is seen a certain number of times (typically, 2-5), and each internal node has appeared more often as a context. A heuristic rule is used for estimating $\hat p$ for each context length, and the optimal context length is chosen such that it will most likely lead to the smallest $\hat h$. Details of this algorithm (which resembles those discussed in Refs. \cite{BCW} and \cite{WRF}) is given in \cite{SG}. 

The information needed to predict a letter with this algorithm consists, on the one hand, of the rules entering the algorithm, and on the other, of the structure stored in the tree. In the present application, we have first build two trees, each based on $\approx 4\times 10^6$ letters from Shakespeare \cite{Sh}, and from the LOB corpus \cite{LOB}, respectively. We have then used this trees to predict additional $\approx 10^6$ letters from these texts. The average estimated entropies were 2.0 bit/character for both texts, which is slightly better than the best published values \cite{BCW}. 

In Figs. 1 and 2, we show the average information per letter as functions of the position in the word \footnote{Technically, a word is defined as any string of letters following a blank and ending with the next blank. Punctuation marks and special characters were taken out in agreement with Ref. \cite{S}, and all non-blank letters were converted to lower case.}. We see indeed a dramatic decrease, both for Shakespeare and for the LOB corpus. The information for the first letter is $\approx 3.8$ bits, which is close to the estimate of 4.1 bit/letter if no contexts are used at all. Thus there is very little information across words which can be used by the algorithm. Already the second letter can be estimated much easier, having an uncertainty of $\approx 2$ bits. This decreased further, until a plateau is reached with the fifth letter where $\hat{\eta}_5\approx 0.7$.

\begin{figure}[ht]
\includegraphics[width=8.0cm,keepaspectratio]{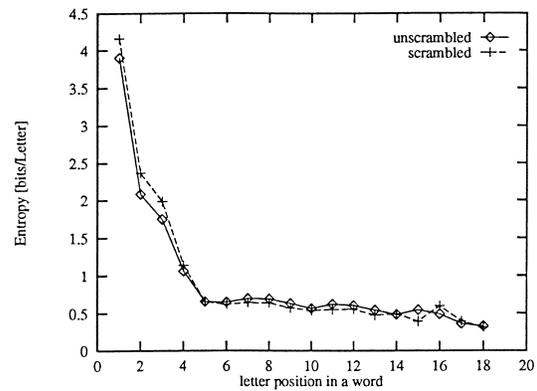} 
\caption{Entropy per letter is dependent on its position in a word for Shakespeare's collected works: Original version ("unscrambled") compared with the surrogate version created by scrambling the words ("scrambled"). Statistics for words longer than 18 letters is too poor to give meaningful estimates.} 
\end{figure}

Actually, we have to be careful when concluding that little information across words can be used by our algorithm. It might be that information is useful for predicting subsequent letters even if it could not be used to predict the first one. To test this, we have created surrogate texts by scrambling the words: all words are permuted randomly, such that any correlation between them is lost while the correlations within words and frequencies of words are unchanged. It increases the average entropies for both text by $\approx 0.1$ bit/letter. The changes in the position-dependent entropies are shown in Figs. 1 and 2. We see that the entropies of the leading letter are increased significantly by scrambling, while those at positions $>4$ are hardly changed at all.

\begin{figure}[ht]
\includegraphics[width=8.0cm,keepaspectratio]{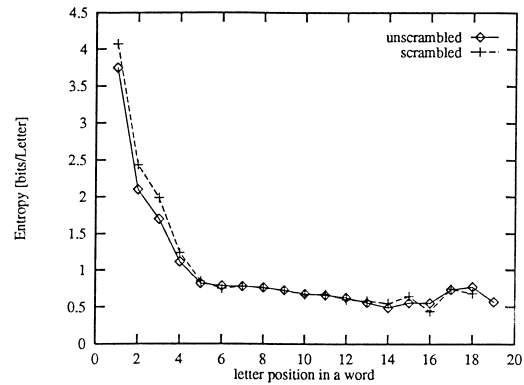} 
\caption{Entropy per letter is dependent on its position in a word for mixed texts from newspapers (LOB corpus): Original version ("unscrambled") compared with the surrogate version created by scrambling the words ("scrambled"). Again, the curves are truncated when the statistical error becomes too large.} 
\end{figure}

Finally, we show in Figs. 3 and 4 how the estimated overall entropy depends on the length of the text, with and without scrambling. That these estimates decrease with the length is a simple consequence of the fact that the algorithm has to "learn" (by building the tree) before being able to make good estimates of $p$. The curves for the scrambled texts are more smooth since the text has been made homogeneous by scrambling. Thus, all learned features will be useful for the future, while this is not true for the unscrambled texts: each time the subject changes, part of the learned features become useless, and new features have to be learned. Thus the convergence of $\hat{h}_N$ for scrambled texts reflects only the learning speed of the algorithm, while that for the unscrambled texts depends also on long range correlations which can be detected only with higher statistics. Extrapolating $\hat{h}_N$ to $N\to\infty$ for unscrambled texts is thus highly non-trivial, as is suggested also by the very low entropies found in \cite{S}-\cite{LR}. In contrast, extrapolation of the curves for scrambled texts is much more easy, and suggests that our estimates for $N\approx 4\times 10^6$ are already very close to the asymptotic ones.

\begin{figure}[ht]
\includegraphics[width=8.0cm,keepaspectratio]{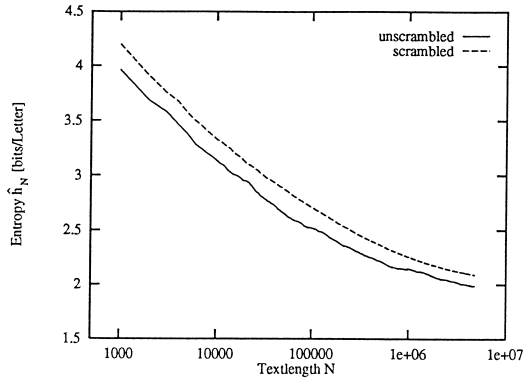} 
\caption{Entropy estimates of Shakespeare's collected works: Original version ("unscrambled") compared with a surrogate version created by scrambling the words ("scrambled").} 
\end{figure}

\begin{figure}[ht]
\includegraphics[width=8.0cm,keepaspectratio]{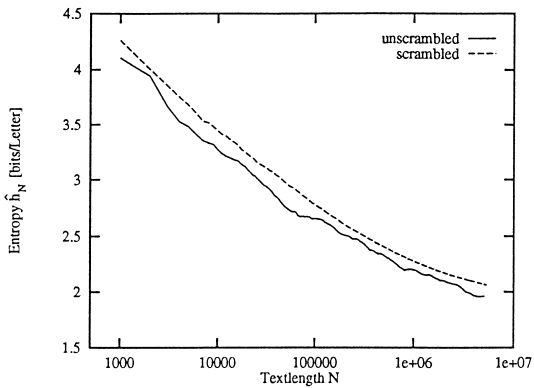} 
\caption{Entropy estimates of mixed English texts from newspapers (LOB corpus): Original version("unscrambled"); surrogate version by scrambling the words ("scrambled").} 
\end{figure}

In summary, we have shown that there are very strong differences in predictability of letters, depending on their position within words. Although such dependencies are to be expected qualitatively, we find the size of the effect surprising. If our algorithm were optimal, it would mean that the constraints within words are indeed much stronger then those between words. But the fact that subjective (human-based) entropy estimates \cite{S}-\cite{LR} are typically lower than machine-based ones, suggest that our algorithm might not be perfect, even though it compares favorably with other algorithms available at present. Thus, our result might just mean that it is harder for the algorithm to learn grammatical (inter-word) than orthographic (intra-word) rules. But in that case, no algorithm of the type used here or in Refs. \cite{BCW} and \cite{WRF} could learn these rules even with much higher computable efforts. Thus, our findings indeed represent an inherent feature of written English, as suggested also by the analysis of scrambled texts. 

Up to now, we have only studied the most primitive grammatical aspects. We should expect similar but less strong differences with the position in a phrase. Other features leading to similar effects could be dependent clauses or direct speech. Obviously, this is a rich field where much remains to be done. Eventually, this could then be used to create more efficient text compression algorithms.\\

This work was supported by DFG within the Graduiertenkolleg "Feldtheoretische und numerische Methoden in der Elementarteilchen- und Statistischen Physik".

\newpage{}
\end{document}